\documentclass[doublecol]{epl2}
\title{Lorentz invariance of entanglement classes in multipartite systems}
\author{M. Huber\inst{1}\thanks{E-mail: \email{marcus.huber@univie.ac.at}} \and N. Friis\inst{2}\thanks{E-mail: \email{pmxnf@nottingham.ac.uk}} \and A. Gabriel\inst{1}\thanks{E-mail: \email{andreas.gabriel@univie.ac.at}} \and C. Spengler\inst{1}\thanks{E-mail: \email{christoph.spengler@univie.ac.at}} \and B. C. Hiesmayr\inst{1,3}\thanks{E-mail: \email{beatrix.hiesmayr@univie.ac.at}}}
\shortauthor{Huber, Friis, Gabriel, Spengler and Hiesmayr}

\institute{
  \inst{1} University of Vienna, Faculty of Physics,  Boltzmanngasse 5, 1090 Vienna, Austria\\
  \inst{2} School of Mathematical Sciences, University of Nottingham, University Park,
Nottingham NG7 2RD, United Kingdom\\
\inst{3} Research Center for Quantum Information, Institute of Physics, Slovak Academy of Sciences, Dubravska cesta 9, 84511 Bratislava, Slovakia
}

\abstract{
We analyze multipartite entanglement in systems of spin-$\frac{1}{2}$ particles from a relativistic perspective. General conditions which have to be met for any classification of multipartite entanglement to be Lorentz invariant are derived, which contributes to a physical understanding of entanglement classification. We show that quantum information in a relativistic setting requires the partition of the Hilbert space into particles to be taken seriously. Furthermore, we study exemplary cases and show how the spin and momentum entanglement transforms relativistically in a multipartite setting.}

\usepackage{amssymb}
\usepackage{graphicx}
\usepackage{amsmath}
\usepackage{amsbsy}
\usepackage{amsthm}
\usepackage{bbm}
\usepackage{bm}
\usepackage{epsfig}
\usepackage{epstopdf}
\usepackage{hyperref}

\newcommand{\be}{\begin{equation}}
\newcommand{\ee}{\end{equation}}
\newcommand{\beq}{\begin{eqnarray}}
\newcommand{\eeq}{\end{eqnarray}}

\usepackage{geometry}
\begin{document}

\maketitle

\begin{figure}[t]
\centering
\includegraphics[scale=2]{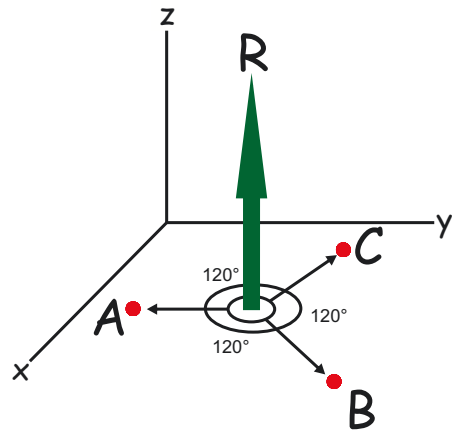}
\caption{Scheme of particle and observer motion in the reference frame
of Alice, Bob, and Charlie}\label{rob}
\end{figure}

\begin{figure}
\centering
\includegraphics[scale=.4]{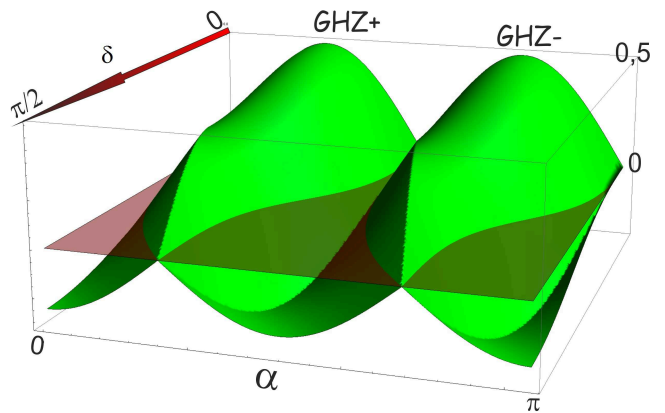}
\caption{Illustration of the violation of inequality~(\ref{multiineq}) from Ref.~\cite{HMGH1} for the state $\sum_i \frac{(-1)^{i}}{\sqrt{6}}|\Pi_i(p_Ap_Bp_C)\rangle\otimes\Bigl(\cos(\alpha)$\mbox{$|\!\downarrow\downarrow\downarrow\rangle$}$+\sin(\alpha)|\!\uparrow\uparrow\uparrow\rangle\Bigr)$. This inequality detects genuine multipartite entanglement and is maximally violated by the GHZ state with an arbitrary phase $\alpha$. This can be seen for the two points maximally violating the inequality, which correspond to the standard GHZ state and the same state with an extra relative phase of $\pi$. The plane corresponds to equality and the area above this plane is detected to be genuinely multipartite entangled. The absolute value of violation of this inequality gives a lower bound on a measure of genuine multipartite entanglement, which is tight for all pure GHZ states (for details see Ref.~\cite{crazychin}). This figure visualizes how the amount of genuine multipartite entanglement decreases as a function of $\delta$.}\label{multiegg}
\end{figure}

In quantum many-body systems multipartite entanglement is a key feature. It plays a central role in a broad variety of physical processes and occurs in various physical systems. In quantum information processing it facilitates quantum computation ({\it e.g.,} Ref.~\cite{qc}), enables multiparty cryptography ({\it e.g.,} Refs.~\cite{SHH1,gisin-crypt}) and quantum algorithms ({\it e.g.,} Ref.~\cite{brussqa}). It appears in quantum phase transitions ({\it e.g.,} Ref.~\cite{phase}) and ionization procedures ({\it e.g.,} Ref.~\cite{helium}). Recently even biological systems have raised questions as to whether multipartite entanglement might be responsible for their astonishing transport efficiency ({\it e.g.} Refs.~\cite{Caruso,bio}).\\
The structure of multiparticle entanglement is however far more complex than the well studied bipartite case (for that see {\it e.g.} Ref.~\cite{horodeckiqe}). Only recently first tools have been developed to answer the question of whether or not a given multi-body quantum state exhibits multipartite entanglement (see, for example, Refs.~\cite{horodeckicrit,wocjancrit,spinchain1,spinchain2,yucrit,hassancrit,hhk}). Further progress was made on experimentally implementable criteria in multipartite systems, first for many qubits in Refs.~\cite{seevinckcrit,guehnewit,guehnecrit} and then for arbitrary higher dimensional systems in Refs. \cite{HMGH1,GHH1}. These criteria answer the question whether multipartite entanglement is present, but it is known that there is a lot of structure beyond merely being entangled. It was shown that there exist several inequivalent ways in which
multipartite systems can be entangled (see {\it e.g.} Refs.~\cite{brussacin,vdmv,Rigolin,bastin}). Questions concerning the number of such entanglement classes and their classification remain unanswered in general. Furthermore, the possible physical implications of these entanglement classes is not fully understood, and addressing this issue will serve as a motivation for this letter.\\
Before we continue to discuss relativistic entanglement let us first briefly review the formal definition of genuine multipartite entanglement in a finite dimensional Hilbert space. Any $n$-partite pure state that can be written as a tensor product
\begin{equation}
|\Psi_{n}\rangle=|\Psi_{B_1}\rangle\otimes|\Psi_{B_2}\rangle
\end{equation}
with respect to some bipartition $B_1|B_2$ is called biseparable. Pure states that are not biseparable with respect to any bipartition are called genuinely multipartite entangled. For mixed states this generalizes in a straightforward way. Any mixed state that can be decomposed into a convex sum of biseparable pure states is called biseparable. Any non-biseparable mixed state is called genuinely multipartite entangled. For details on the intricacies that mixed state biseparability provides consult e.g. Refs.~\cite{seevinckcrit,guehnewit,guehnecrit,HMGH1,GHH1}. Famous examples of genuinely multipartite entangled states are the GHZ (Greenberger, Horne, Zeilinger) state and the W state which will be used later.\\
Studying entanglement in a relativistic framework takes quantum information one step further. In a relativistic setting many new features appear which are not present in a non-relativistic framework. Although a self-contained description of relativistic quantum information has not yet been formulated, the identification of observer independent quantities ought to be crucial to such an endeavor.
 It has been shown that the entanglement between the spins of two particles is not Lorentz invariant (see {\it e.g.} Refs.~\cite{GingrichAdami,Lamata,relqi,relqi2,relqi3,relqi4,relqi5}). In a more general approach, taking into account not only the spin entanglement, but also the momenta of the relativistic particles, it has recently been shown, that in bipartite systems Lorentz invariance of entanglement can generally only be claimed for the Hilbert space partition into individual particles (for details see Ref.~\cite{FBHH1}). The entanglement in other partitions is observer dependent (i.e. it can be transferred between them).\\
The effect of relativistic entanglement transformation is not limited to only systems of two particles. In this paper we show that in a relativistic multi-particle setting the different classes of multipartite entanglement are indeed observer independent. We also provide a general framework to identify which conditions have to be met by a possible classification of multipartite entanglement in order to be Lorentz invariant. Furthermore, we show how the
entanglement in different partitions changes through Lorentz transformations,
while it is preserved for the partition into individual particles, which is thus
singled out naturally.\\
We first start with an illustrative example of a genuinely multipartite entangled system of three spin-$\frac{1}{2}$ particles, observed from two different reference frames. We then discuss in detail the effect of Lorentz transformations on the entanglement of this exemplary system. Finally we prove that the results also hold in a general setting.\\
Let Alice, Bob, and Charlie be inertial observes, resting in a common frame of reference, who share a quantum state which is separable between spins and momenta
\begin{eqnarray}
|\Psi_{ABC}\rangle=|\psi_{\text{mom}}\rangle\otimes|\phi_{\text{spin}}\rangle\, ,
\label{exstate}
\end{eqnarray}
where $|\psi_{\text{mom}}\rangle$ is the state of the momenta and $|\phi_{\text{spin}}\rangle$ is the state of the spins. Let us further assume that they share a multipartite entangled spin state, e.g. the well known GHZ state
\begin{eqnarray}
\label{exstates}
|\phi_{\text{spin}}\rangle=\frac{1}{\sqrt{2}}(|\downarrow\downarrow\downarrow\rangle+|\uparrow\uparrow\uparrow\rangle)\, ,
\end{eqnarray}
with which they want to perform a quantum protocol (e.g., quantum secret sharing as in Ref.~\cite{SHH1}). Now a relativistic observer (Robert) is moving perpendicular to the plane of their shared quantum state in the $z$-direction (see Fig.~\ref{rob} for details). From his point of view the situation of course looks different, as he observes a Lorentz-boosted state $\rho_\Lambda$.
Let us also, for the sake of simplicity and without loss of generality (regarding our result) assume that we have momentum eigenstates, i.e. $P^\mu_i|\psi(p_1,p_2,p_3)\rangle=p^\mu_i|\psi(p_1,p_2,p_3)\rangle$ and only three possible momenta. Let our three sharp momenta be denoted by $p_A$, $p_B$ and $p_C$, where $|\vec{p}_A|=|\vec{p}_B|=|\vec{p}_C|$. The Lorentz boost to Robert's frame induces Wigner rotations of the spins, i.e. for a separable momentum state the state (2) is transformed to
\begin{eqnarray}
|\psi(\tilde{p}_1,\tilde{p}_2,\tilde{p}_3)\rangle\otimes(U_{local}(p_1,p_2,p_3,\delta))|\phi_{\text{spin}}\rangle\, ,
\end{eqnarray}
where
\begin{equation}
U_{\text{local}}(p_1,p_2,p_3,\delta)=U(p_1,\delta)\otimes U(p_2,\delta)\otimes U(p_3,\delta)\, ,
\end{equation}
are local unitary operations on the spin vector, and the Wigner rotation angle $\delta$ is given by
\begin{equation}
\tan\delta\,=\,\frac{\sinh\eta\,\sinh\xi}{\cosh\eta\,+\,\cosh\xi}\ \ \,,
\label{eq:wigner rotation angle}
\end{equation}
where $\eta$ and $\xi$ are the rapidities corresponding to the velocities $u$ of Robert and $v$ of the three particles relative to Alice's, Bob's and Charlie's frame of reference, given by $\tanh \eta=u$ and $\tanh \xi=v$ (for more details on Wigner rotations consult e.g. Refs.~\cite{FBHH1,alsingmilburn}). The assumption of sharp momenta, which is a common approximation (see, e.g., Ref.~\cite{relqi}), allows us to apply a single Wigner rotation
for each particle momentum.\\
The entanglement class of the spin state remains invariant, if the momentum state is separable as in this case the spin state undergoes only a local unitary transformation. So this yields the first result on the conditions of entanglement classes in a relativistic setting:\\
\newpage
\textbf{Condition 1}:~
\begin{itemize}
	\item[] Different Lorentz invariant classes of multipartite entanglement need to be inequivalent under local unitary operations.
\end{itemize}
That this is a necessary condition follows simply from the fact that even for separable momentum states the Lorentz boost acts as a local unitary transformation. So if two local unitary equivalent states were in a different entanglement class the class membership would fail to be Lorentz invariant. Also it would be rather meaningless, as any local basis change (e.g. relabeling the measurement apparatuses) could change the classification of an investigated state. This condition is satisfied for all previously introduced entanglement classification schemes. Those are usually defined via SLOCC (stochastic local operations and classical communication)-inequivalent states (see, {\it e.g.}, Ref.~\cite{brussacin,bastin}), which incorporate local unitary operations. 
Now we focus on the case where the momentum state is not separable, e.g.
\begin{equation}
\label{exstatem}
|\psi(p_1,p_2,p_3)\rangle=\sum_i \alpha_i |\Pi_i(p_Ap_Bp_C)\rangle\,,
\end{equation}
where $\Pi_i$ denotes a permutation of $p_Ap_Bp_C$ (an even permutation for even $i$ and an odd permutation for odd $i$) and the sum is taken over all possible permutations. In this case the spin state transforms as
\begin{eqnarray}
\lefteqn{\Lambda[\rho_{\text{spin}}]=\sum_i|\alpha_i|^2\cdot}\\&& U_{\mathrm{local}}(\Pi_i(p_A,p_B,p_C),\delta)\rho_{\text{spin}}U^\dagger_{\mathrm{local}}(\Pi_i(p_A,p_B,p_C),\delta)\, ,\nonumber
\end{eqnarray}
where $\rho_{\text{spin}}=|\phi_{\text{spin}}\rangle\langle\phi_{\text{spin}}|$ and $\Lambda$ denotes the appropriate representation of the Lorentz transformation. Equivalently to the bipartite case the spin state becomes more mixed and entanglement decreases as we have plotted in detail for our example state in Fig.~\ref{multiegg}. In a more general setting
of unsharp momenta, the sum would have to be replaced
by an integral and the coefficients $\alpha_{i}$ would be exchanged
with an appropriate distribution function in momentum
space, (implicitly) containing all additional information
about the state (e.g., position and orbital angular momentum). Now it is evident which further condition has to be met for a classification in order to be observer independent:\\

\textbf{Condition 2}:~\begin{itemize}
	\item[] Any convex combination of local-unitarily equivalent pure states defines a Lorentz invariant class of genuine multipartite entanglement.
\end{itemize}
To prove this condition it is sufficient to look at the most general setting and work out which Wigner rotations the Lorentz boost induces. The most general state in the resting frame may be an arbitrarily mixed state:
\begin{equation}
\rho=\sum_iq_i|\Psi_{\text{mom+spin}}^i\rangle\langle\Psi_{\text{mom+spin}}^i|\, ,
\end{equation}
where $|\Psi_{\text{mom+spin}}^i\rangle=\sum_k\alpha^i_k|\psi^k_{\text{mom}}\rangle_i\otimes|\phi_{\text{spin}}^k\rangle_i$.
In this case the state of the spins is given by
\begin{eqnarray}
\rho_{\text{spin}}&=&\sum_iq_i\text{Tr}_{\text{mom}}(|\Psi_{\text{mom+spin}}^i\rangle\langle\Psi_{\text{mom+spin}}^i|)\nonumber\\
&=&\sum_iq_i\sum_{k}|\alpha^i_k|^2\underbrace{|\phi_{\text{spin}}^k\rangle_i\langle\phi_{\text{spin}}^k|_i}_{\sigma_i^k}\, .
\end{eqnarray}
\\
If we look at the state from a perspective from which a Lorentz boost induces a Wigner rotation, the reduced spin state is given as
\begin{eqnarray}
\lefteqn{\Lambda[\rho_{\text{spin}}]=\sum_iq_i\sum_{k}|\alpha^i_k|^2\cdot}\\&&{U^k}_{\text{local}}(p_1,p_2,\cdots, p_n,\delta)\sigma_i^k{U^k}^\dagger_{\text{local}}(p_1,p_2,\cdots, p_n,\delta)\, .\nonumber
\end{eqnarray}
Certainly, if the initial spin states $|\phi_{\text{spin}}^k\rangle_i$ lies within a certain equivalence class $\mathcal{C}$, then their convex
combination $\rho_{\text{spin}} \in \mathcal{C}$ as well. Defining an entanglement class by Condition
1 it is evident that also $\Lambda[\rho_{\text{spin}}] \in \mathcal{C}$, since all $U^k_{\text{local}} \sigma_i^k
{U^k_{\text{local}}}^\dagger$ must remain within $\mathcal{C}$, thus proving that Condition 2 is
necessary and sufficient for a class of genuine multipartite entangled states to
be observer independent. In the more general case of unsharp momenta the statement remains true, as the Wigner rotations are only induced by the momentum part of the system. In this case the discrete sum would change into a continuous integral over such local unitarily rotated states. This would surely make the analysis harder (although with the methods introduced in Ref.~\cite{SHH3} it can even be done analytically in arbitrary dimensions), but the sufficiency of the condition of course remains unchanged.\\
\begin{figure}[t!]
\centering
(a)\includegraphics[scale=.65]{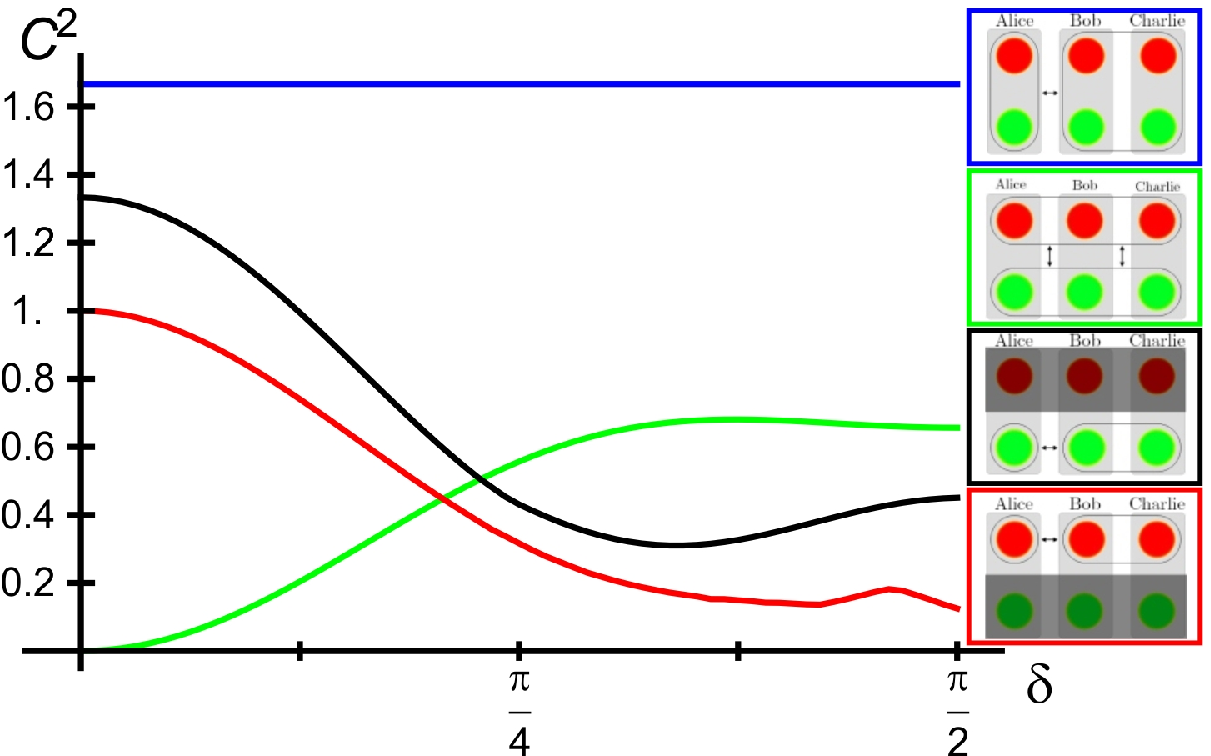}\\
(b)\includegraphics[scale=.65]{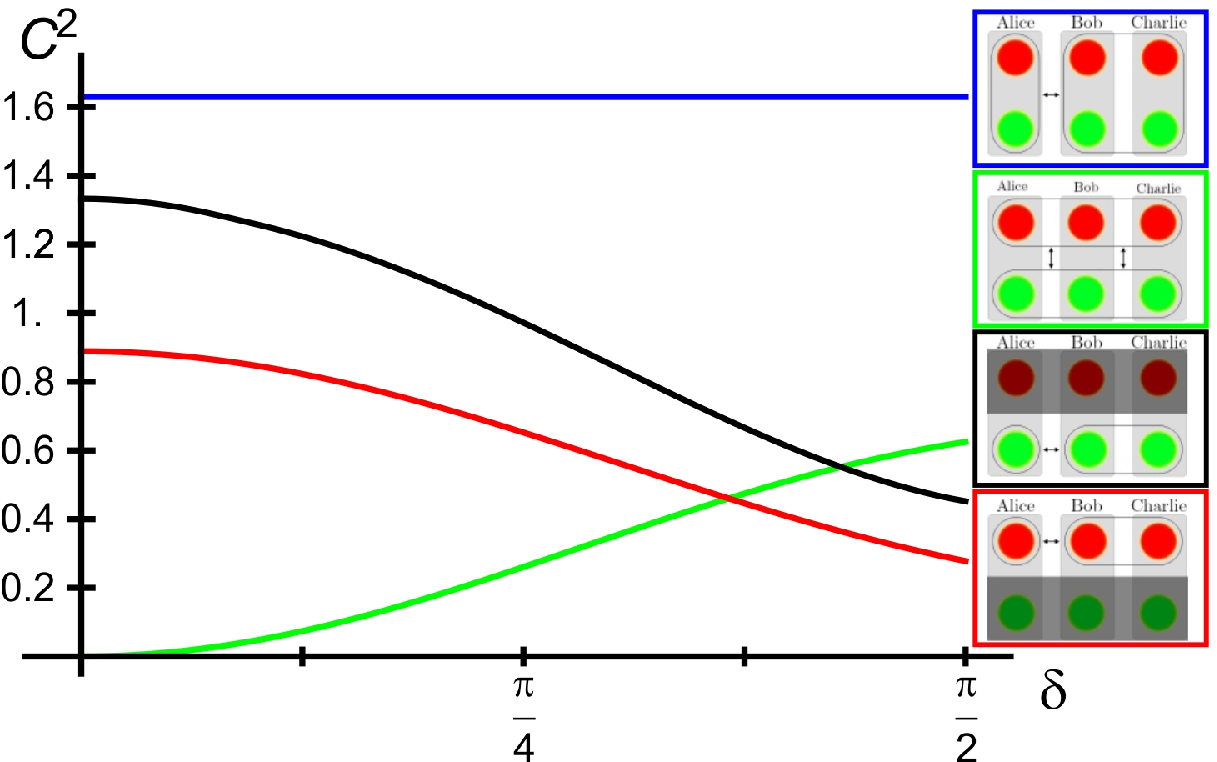}

\caption{Illustration of how the entanglement for different partitions transforms depending on $\delta$. To measure entanglement we used the m-concurrence introduced in Refs.~\cite{HH2,hhk} and optimized it using the approach from Ref.~\cite{SHH2}. In (a) the entanglement of exemplary state (\ref{exstate}) with momentum state (\ref{exstatem}), where $\alpha_i=\frac{(-1)^i}{\sqrt{6}}$, and spin state (\ref{exstates}) is plotted and in (b) the spin state is replaced by a W state $|W\rangle=\frac{1}{\sqrt{3}}(|\!\downarrow\downarrow\uparrow\rangle+|\!\downarrow\uparrow\downarrow\rangle+|\!\uparrow\downarrow\downarrow\rangle)$. Each graph corresponds to the partition depicted in the same order on the right, where the top (red) dots represent the spins and the bottom (green) dots represent the momenta of the three particles.}\label{partition}
\end{figure}
The resulting conditions do not unambiguously determine the classification of multipartite entanglement. They are however a necessary criterion for physical consistency of any possible classes. Indeed the previously defined entanglement classes from Ref.~\cite{brussacin} and the more general SLOCC classification from Ref.~\cite{bastin} meet this condition. 
In Ref.~\cite{HSGSHB1} the authors introduce an experimentally accessible classification scheme of multiqubit entanglement, which incorporates the one introduced in Ref.~\cite{brussacin}, and is indeed also observer independent.\\
In our exemplary figures we illustrate how the amount of entanglement changes through the Lorentz transformation. In Fig.~\ref{multiegg} we plot a lower bound of a measure of genuine multipartite entanglement, which is tight in the case of GHZ states. It is based on the non-linear entanglement witness introduced in Ref.~\cite{HMGH1} and can be experimentally ascertained using only few local measurements (in the case of three qubits only $9$ local measurement settings, as opposed to a full state tomography requiring $63$). That it also serves as a lower bound to an entanglement measure quantifying genuine multipartite entanglement was shown in Ref.~\cite{crazychin}.\\
While the classification itself does not depend on any observer's knowledge of the entire state, Robert's ability to unambiguously determine the class of the reduced density matrix of the spins may. Whereas the class remains invariant for all observers, the separability properties may change, as depicted in Fig.~\ref{multiegg} and Fig.~\ref{partition}~(a) and (b). In particular, if the momentum state is entangled, the reduced spin density matrix of a previously genuinely multipartite entangled state may become biseparable through the Lorentz transformation. As long as the initial state is separable with respect to spin and momentum, the entanglement of the reduced density matrix of the spins can never increase (only decrease in accordance to the main theorem of Ref.~\cite{GingrichAdami}). Since biseparable states can result from convex mixtures of any class of genuinely multipartite entangled states, Robert necessarily needs to also have information about the momentum state (which allows for an unambiguous decomposition) in order to determine the class unambiguously in this case.\\

In conclusion we have shown which conditions have to be met for an entanglement
classification scheme to be Lorentz invariant. We have further argued why
knowledge of the momentum state is helpful to a complete classification of
genuine multipartite entanglement of the spin state, which is not surprising, as
it is well known that the reduced spin density matrix does not transform
covariantly under Lorentz boosts. Nonetheless the proposed classification
scheme is retaining its Lorentz invariance if the momentum state is unknown, i.e.
all inertial observers will assign a given state to an entanglement class or, at
most, to one of the corresponding convex subsets of this class. This provides a
general framework which paves the way for entanglement classification beyond $n$ qubits and, at the same time, imposes an intuitive physical understanding upon
the distinction into different entanglement classes.

\acknowledgments
We would like to thank R.~A.~Bertlmann, F.~Hipp, S.~Radic, H.~Schimpf and T.~Adaktylos for productive discussions.
M.~H., A.~G. and C.~S. gratefully acknowledge the Austrian Fund project FWF-P21947N16. A.~G. is supported by the University of Vienna's research grant. N.~F. acknowledges support from EPSRC [CAF Grant No.~EP/G00496X/2 to I.Fuentes]. B.~C.~H. acknowleges the EU project QESSENCE.\\

\hspace*{0.3cm}APPENDIX\\

The explicit form of inequality~(\ref{multiineq}) from Ref.~\cite{HMGH1} written in terms of local spin observables for three spin-$\frac{1}{2}$ systems reads

\begin{align}
\label{multiineq}
&   \left|\frac{1}{4}\langle\sigma_x\sigma_x\sigma_x-\sigma_x\sigma_y\sigma_y-\sigma_y\sigma_x\sigma_y-\sigma_y\sigma_y\sigma_x\rangle\right.   \nonumber\\
&   \,\left.+\frac{1}{4}\langle\sigma_y\sigma_y\sigma_y-\sigma_x\sigma_x\sigma_y-\sigma_y\sigma_x\sigma_x-\sigma_x\sigma_y\sigma_x\rangle\right|  \nonumber\\
&   \,-\sqrt{\langle P_1^+P_2^+P_3^-\rangle\langle P_1^-P_2^-P_3^+\rangle}    \nonumber\\
&   \,-\sqrt{\langle P_1^+P_2^-P_3^+\rangle\langle P_1^-P_2^+P_3^-\rangle}    \nonumber\\
&   \,-\sqrt{\langle P_1^-P_2^+P_3^+\rangle\langle P_1^-P_2^+P_3^+\rangle}    \tag{A.1}
\end{align}
where
\begin{align}
P^\pm := \frac{(1\pm\sigma_z)}{2}.\tag{A.2}
\end{align}

%
%
%
%

\end{document}